\documentclass[twocolumn]{aastex63}

\graphicspath{{./}{figures/}}
\usepackage{natbib}

\begin{document}

\title{Global X-Ray Properties of the Vela and Puppis A Supernova Remnants}

\author{E.M. Silich}
\affil{Department of Physics and Astronomy, University of Iowa, Iowa City, IA 52242, USA}

\author{P. Kaaret}
\affil{Department of Physics and Astronomy, University of Iowa, Iowa City, IA 52242, USA}

\author{A. Zajczyk}
\affil{Department of Physics and Astronomy, University of Iowa, Iowa City, IA 52242, USA}
\affil{NASA Goddard Space Flight Center, Greenbelt, MD 20771, USA}
\affil{Center for Space Sciences and Technology, University of Maryland, Baltimore County, 1000 Hilltop Circle, Baltimore, MD 21250}

\author{D.M. LaRocca}
\affil{Department of Physics and Astronomy, University of Iowa, Iowa City, IA 52242, USA}

\author{J. Bluem}
\affil{Department of Physics and Astronomy, University of Iowa, Iowa City, IA 52242, USA}

\author{R. Ringuette}
\affil{Department of Physics and Astronomy, University of Iowa, Iowa City, IA 52242, USA}

\author{K. Jahoda}
\affil{NASA Goddard Space Flight Center, Greenbelt, MD 20771, USA}

\author{K. D. Kuntz}
\affil{The Henry A. Rowland Department of Physics and Astronomy, Johns Hopkins University, Baltimore, MD 21218, USA}

\begin{abstract}
The Vela and Puppis A supernova remnants (SNRs) comprise a large emission region of $\sim 8^{\circ}$ diameter in the soft X-ray sky. The HaloSat CubeSat mission provides the first soft X-ray ($0.4-7$ keV) observation of the entire Vela SNR and Puppis A SNR region with a single pointing and moderate spectral resolution. HaloSat observations of the Vela SNR are best fit with a two-temperature thermal plasma model consisting of a cooler component with $kT_{1} = 0.19^{+0.01}_{-0.01}$ keV in collisional ionization equilibrium and a hotter component with $kT_{2} = 1.06^{+0.45}_{-0.27}$ keV in non-equilibrium ionization. Observations of the Puppis A SNR are best fit with a single-component plane-parallel shocked plasma model with $kT = 0.86^{+0.06}_{-0.05}$ keV in non-equilibrium ionization. For the first time, we find the total X-ray luminosities of both components of the Vela SNR spectrum in the $0.5-7$ keV energy band to be $L_X = 4.4^{+1.4}_{-1.4} \times 10^{34}$ erg s$^{-1}$ for the cooler component and $L_X = 4.1^{+1.8}_{-1.5} \times 10^{34}$ erg s$^{-1}$ for the hotter component. We find the total X-ray luminosities of the Vela and Puppis A SNRs to be $L_{\text{X}} = 8.4 \times 10^{34}$ erg s$^{-1}$ and $L_X = 6.7^{+1.1}_{-0.9} \times 10^{36}$ erg s$^{-1}$.
\end{abstract}

\section{Introduction} \label{sec:intro}

The Vela supernova remnant (SNR) is among the brightest and largest sources in the soft X-ray sky and has been used to construct global models of the dynamics of SNRs \citep{2011A&A...525A.154S}. It is estimated to be a middle-aged SNR at around 11.4 kyr old \citep{1999ApJ...515L..25C} at a distance of $250\pm30$ pc \citep{1999ApJ...515L..25C} with its primary X-ray emission due to two thermal components of heated interstellar cloud matter \citep{2000A&A...362.1083L}. Located at the coordinates $\alpha$ (J2000) = $8^{\text{h}} \: 35^{\text{m}} \: 20.7^{\text{s}}$, $\delta$ (J2000) = $-45^{\circ} \: 10' \: 35.7''$, the Vela SNR is superimposed upon the Puppis A SNR, located at coordinates $\alpha$ (J2000) = $8^{\text{h}} \: 21^{\text{m}} \: 56.7^{\text{s}}$, $\delta$ (J2000) = $-43^{\circ} \: 00' \: 19.0''$ (see Fig. 1), which is also among the brightest sources in the soft X-ray sky. Younger than the Vela SNR at an estimated age between 3.7 and 4.45 kyr \citep{1988srim.conf...65W, 2012ApJ...755..141B}, the Puppis A SNR is believed to be at a distance of 2.2 kpc \citep{2003MNRAS.345..671R} with an X-ray spectrum dominated by shock-heated interstellar material \citep{2008ApJ...676..378H}. \par

\citet{2000A&A...362.1083L} performed a spatially resolved X-ray spectral analysis of the entire $\sim8^{\circ}$ extent of the Vela SNR by fitting individual spectra across the SNR with ROSAT data.  They found the spectra to be consistent with a two-temperature Raymond-Smith thermal plasma model in CIE with asymmetric temperatures, emission measures, and interstellar absorption column densities across the SNR in the $0.1-2.5$ keV band. However, ROSAT has inadequate spectral resolution to reveal potential non-equilibrium ionization (NEI) spectral contributions, so NEI models could not be distinguished from CIE models in their analysis.

A comprehensive survey of the entire Vela SNR region with moderate (CCD-like) spectral resolution has not been previously done due to the large size of the remnant and the typically small field of view (FOV) of current X-ray missions. Global measures of SNR emission, particularly those resolved into different temperature components, are required to model the overall dynamics of the remnant; \citet{2011A&A...525A.154S} noted that the lack of a luminosity for the higher temperature component placed an uncertainty on the density of the hot component, which they then attempted to obtain through other methods. HaloSat provides this measurement at the CCD energy resolution typical of current missions but with a superior signal-to-noise ratio.

HaloSat is a CubeSat mission which is sensitive to soft X-ray emission in the $0.4 - 7$ keV band \citep{2019ApJ...884..162K}. It uses three independent non-imaging silicon drift detectors, each of which has a FOV with 5$^{\circ}$ radius full response that falls to zero response at a radius of 7$^{\circ}$. HaloSat observations of the Vela SNR are unique in that they are able to contain the
entire soft X-ray emission feature in a single FOV with moderate spectral resolution (see Fig. 2). This allows global properties such as flux, luminosity, temperature, and ionization state of each SNR to be determined. In order to disentangle the spectral contributions from each SNR, two HaloSat pointings are used, one of which encompasses the entirety of the Vela and Puppis A SNRs, and the other is offset so that the Puppis A SNR contributions are maximized while the Vela SNR contributions are minimized. \par

\begin{figure}[t]
\epsscale{1.25}
\plotone{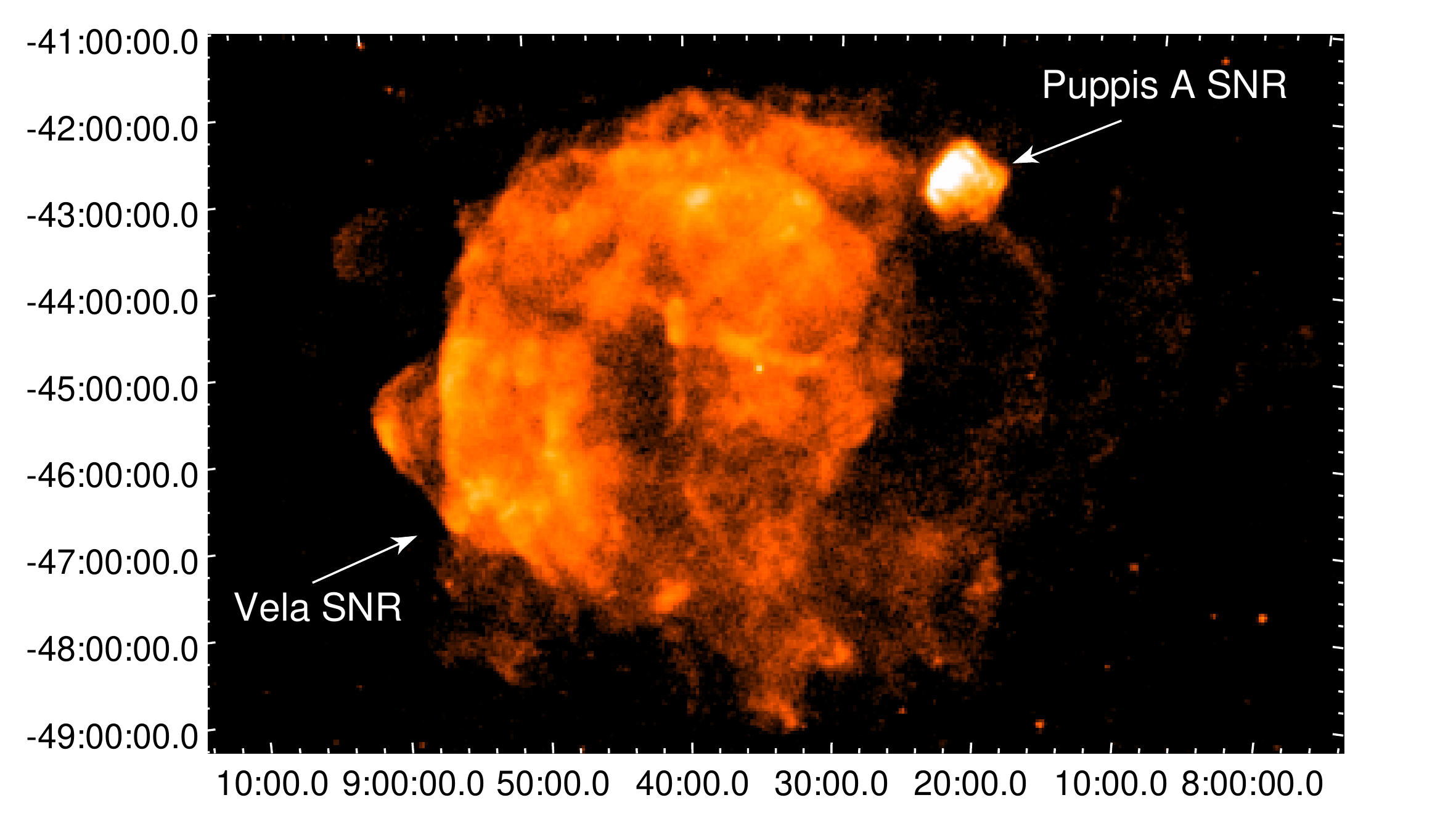}
\caption{The Vela SNR and Puppis A SNR indicated on a ROSAT $0.1-2.04$ keV map.}
\end{figure}

We describe HaloSat observations of the Vela SNR and Puppis A SNR region. Observations and data reduction methods are given in section 2. Spectral analysis procedures and results are described in section 3. Our results, including the best-fit spectral model, X-ray fluxes and luminosities of the Vela SNR and Puppis A SNR, and a comparison to previous observations are detailed in section 4. Conclusions are given in section 5.

\section{Observations and Data Reduction} \label{sec:obs}

\begin{figure}[b]
\epsscale{1.35}
\plotone{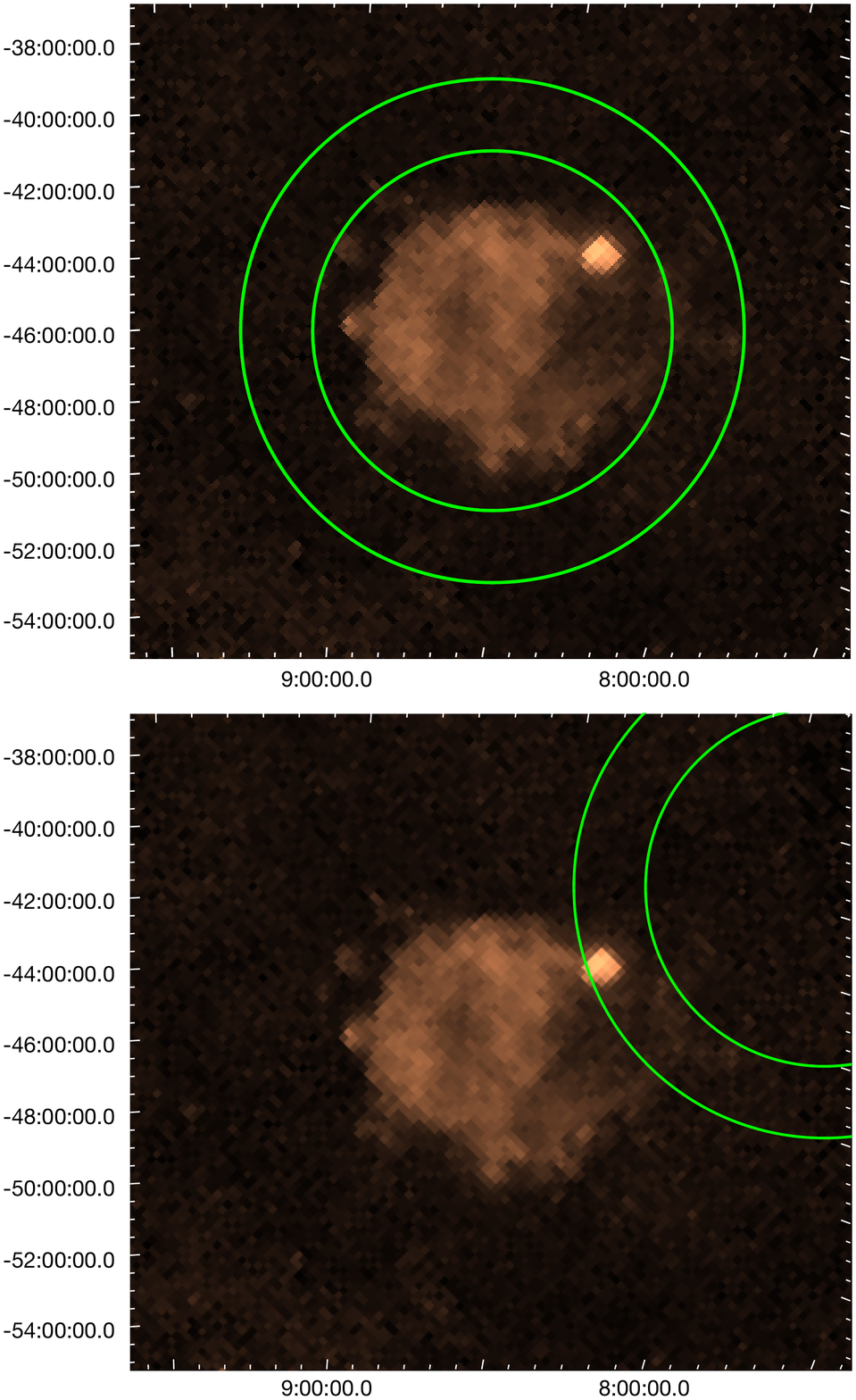}
\caption{HaloSat target pointings indicated on a ROSAT \textbf{$0.56-1.21$} keV map. Green inner circles represent the $5^{\circ}$ radius full response FOV, and the outer circles represent the $7^{\circ}$ radius zero response FOV for each pointing. \textit{Top}: HS0007 pointing centered about the Vela SNR. \textit{Bottom}: HS0009 pointing offset from both the Vela and Puppis A SNRs. }
\end{figure}

\begin{deluxetable*}{ccccc}[t]
\tablecaption{HaloSat Observations of the Vela SNR and Puppis A SNR Region \label{tab:table}}
\tablehead{
\colhead{Target Name} & \colhead{ID} & \colhead{RA} & \colhead{Dec} & \colhead{Dates Observed}}
\startdata
Vela SNR & HS0007 & $8^{\text{h}} \: 37^{\text{m}} \: 9.1^{\text{s}}$ & $-45^{\circ} \: 30' \: 43.2''$ & 2018 Dec $11 - 2019$ May 9 \\
Puppis A SNR Offset & HS0009 & $7^{\text{h}} \: 52^{\text{m}} \: 18^{\text{s}}$ & $-39^{\circ} \: 19' \: 51.6''$ & 2018 Dec $10 - 2019$ Feb 14 \\
\enddata
\end{deluxetable*}

HaloSat observations of the Vela SNR and Puppis A SNR region consist of two instrument pointings (see Table 1, Fig. 2). The first pointing (HS0007) is centered about the Vela SNR at coordinates $(\alpha, \delta) = (129.288^{\circ}, -45.512^{\circ})$ (J2000). The goal of HS0007 is to observe both the Vela SNR and Puppis A SNR within the 5$^{\circ}$ radius full response FOV with a single observation. The second pointing (HS0009) is observed at coordinates $(\alpha, \delta) = (118.075^{\circ}, -39.331^{\circ})$ (J2000). It is offset from both the Vela SNR and Puppis A SNR in order to maximize the total emission received from the Puppis A SNR while minimizing that from the Vela SNR. This allows for the spectral contributions from the Puppis A SNR to be effectively isolated from those of the Vela SNR, so that the individual source spectra can be modeled accurately. A cut is applied to the data to restrict the spacecraft pointing offset to within 0.01$^{\circ}$ of the intended coordinates for each target. The individual detectors may be offset by up to $\pm 0.23^{\circ}$ from the pointing direction for each target \citep{2019ApJ...884..162K}. \par

 Data are processed in order to select Good Time Intervals (GTIs) with acceptable background contributions. Cuts are applied to the data on specified energy bands. The $3-7$ keV energy band is dominated by the time-variable instrumental particle background, and events above 7 keV are primarily due to large depositions of charge in the detectors by high energy X-rays or charged particles. Cuts are applied to restrict the count rate in the $3-7$ keV energy band in 64 s time bins to $\leq 0.20$ counts sec$^{-1}$ to reduce the time-variable background contributions. The count rate of very large events (VLEs) above 7 keV in 64 s time bins is restricted to $\leq 1.875$ counts sec$^{-1}$. The detectors contribute data in the entire $0.4-7.0$ keV band, but one of the detectors (DPU 14) produced data that was in poor agreement with the other two detectors at low energies, so for DPU 14, only data in the $1-7$ keV band were used. After filtering, HS0007 has a total combined exposure of 133.4 ks, and HS0009 has a total combined exposure of 80.1 ks for all detectors. The resulting spectra are binned with the GRPPHA tool from FTOOLS in order to obtain a minimum of 40 counts per bin to allow use of chi-squared statistics in the fitting. \par

 The HaloSat on-orbit instrument calibration is described by \citet{2019ApJ...884..162K}. Spectra are formatted to the OGIP standards suitable for XSPEC analysis, and the response matrix is composed of the response files that are current as of 2019 April 23.

\section{Spectral Analysis}  \label{sec:analysis}

Spectra are simultaneously fit using the XSPEC X-ray spectral fitting package (version 12.10.1f). The cosmic X-ray background (CXB) contributions are modeled with a single absorbed power law with $\Gamma = 1.45$ and a 1 keV normalization of 10.91 keV cm$^{-2}$ s$^{-1}$ sr$^{-1}$ keV$^{-1}$ \citep{2017ApJ...837...19C}. Contributions from the local hot bubble (LHB) are modeled using an unabsorbed \textsc{apec} model with $kT = 0.11$ keV and an emission measure of 0.0034 cm$^{-6}$ pc along the Vela SNR line of sight found by \citet{2017ApJ...834...33L}. \par
The HaloSat instrumental background contributions are modeled with a single power law with a fixed photon index for each detector (DPU) and a free normalization for each DPU for each target. The fixed photon indices are calculated through analysis of the HaloSat background for a large set of halo observations with Galactic latitudes $|b| > 30^{\circ}$. The instrumental background dominates at energies above 2 keV (see Fig. 3), and it appears to be well described by a power-law in that interval, similar to the background observed for \textit{XMM-Newton}. The low energy extension of the background powerlaw was tested by applying different sets of filtering cuts for the halo fields. Consistent results were obtained for the halo emission on each field. The emission from Vela and Puppis A in the $0.5-1.5$ keV band, of primary interest for the results obtained below, is well above the instrument background. Photon indices and corresponding fitted normalizations are given in Table 2. \par
Other known sources along the Vela SNR line of sight include the Vela Pulsar, Vela X pulsar wind nebula (PWN), Vela Jr. SNR, and PSR J0855-4644 with its PWN. The combined contributions from the $0.2 - 2$ keV flux of the Vela Pulsar \citep{2001ApJ...552L.129P}, the $0.3 - 7$ keV flux of the Vela X PWN \citep{2010ApJ...713..146A}, the $0.1 - 2.4$ keV flux of the Vela Jr. SNR \citep{1998Natur.396..141A}, and the $2 - 10$ keV flux of SR J0855-4644 and its PWN \citep{2013A&A...551A...7A} account for less than $0.75\%$ of the total X-ray flux of the Vela and Puppis A SNRs in the $0.1 - 2.5$ energy range of interest. Total flux contributions from X-ray transient sources in the MAXI GSC Monitoring Results (v6l) \citep{2009PASJ...61..999M} and point sources from the ROSAT All-Sky Survey Bright Source Catalogue \citep{1999A&A...349..389V} are within the statistical uncertainty of our flux measurements. Thus, these sources do not have a significant effect on our results. \par

	The HaloSat spectra from HS0007 and HS0009 contain emission from both the Vela and Puppis A SNRs. In order to model this, we use one set of physical source parameters for the Vela SNR emission and one set for the Puppis A SNR emission. The physical source parameters are then linked so that the same set of parameters for each source is used when fitting the two spectra simultaneously in XSPEC. Another parameter we define is the response-weighting factor, which does not relate to the physical source parameters, but accounts for how the HaloSat observations were designed and executed.
	Since HS0007 observes both the Vela SNR and Puppis A SNR in the central $5^{\circ}$ radius of the FOV that has full response and HS0009 excludes most of the Vela SNR and includes almost all of the Puppis A SNR in the region between $5^{\circ}$ and $7^{\circ}$ in the FOV where the HaloSat response decreases linearly from full-response to zero-response, the detector count rate in HS0009 is reduced relative to that of HS0007. In the fitting procedure, the response-weighting factors are defined so that the fraction of the Vela SNR flux observed in the HS0009 spectrum is the `Vela response-weighting factor', and the `Puppis A response-weighting factor' is the ratio of the total detector count rate for the Puppis A SNR in HS0007 to that in HS0009. The inverse of the Puppis A response-weighting factor is the fractional reduction in count rate for HS0009 relative to HS0007.

\begin{deluxetable}{cccc}[t]
\tablecaption{Instrumental background parameters for the best-fit model \label{tab:table}}
\tablehead{
\colhead{Target ID} & \colhead{DPU ID} & \colhead{$\Gamma$} & \colhead{Normalization}}
\startdata
HS0007 & 14 & 0.651 & $0.031^{+0.001}_{-0.002}$ \\
HS0007 & 54 & 0.593 & $0.027^{+0.001}_{-0.002}$ \\
HS0007 & 38 & 0.655 & $0.029^{+0.001}_{-0.002}$ \\
HS0009 & 14 & 0.651 & $0.025^{+0.001}_{-0.001}$ \\
HS0009 & 54 & 0.593 & $0.021^{+0.001}_{-0.001}$ \\
HS0009 & 38 & 0.655 & $0.023^{+0.001}_{-0.001}$
\enddata
\end{deluxetable}

The Puppis A spectral contributions are modeled with an absorbed (\textsc{tbabs}) single-component plane-parallel shocked plasma in non-equilibrium ionization (\textsc{pshock}; \citet{2001ApJ...548..820B}) model using AtomDB data with \textsc{wilm} \citep{2000ApJ...542..914W} abundances set to solar values. The \textsc{pshock} model is used to represent NEI thermal plasma components during analysis due to its common usage in modeling the Puppis A SNR soft X-ray spectrum \citep{2008ApJ...676..378H, 2012ApJ...756...49K, 2016A&A...590A..70L}. \citet{2016A&A...590A..70L} notes that larger regions of the Puppis A SNR may be better fit with the two-temperature shocked plasma model, as opposed to the single-temperature model, which produces acceptable fits on smaller scales. As the HaloSat observations include the entire SNR, the two-temperature model was tested in analysis, but it did not produce a significantly better fit to the data than the single-component model, with an FTEST probability of $0.52$. Thus, for simplicity the single-component \textsc{pshock} model was adopted to represent the Puppis A SNR spectrum. Free parameters include the interstellar absorption column density, temperatures, upper ionization timescales, and normalizations \citep{2016A&A...590A..70L, 2008ApJ...676..378H}.\par

Three separate models are tested in order to determine whether a CIE model (\textsc{apec}; \citet{2001ApJ...556L..91S}), an NEI model (\textsc{pshock}), or a combination of both is better representative of the Vela SNR thermal spectral components. Each model of the Vela SNR utilizes two different temperature components, where the interstellar absorption column density, plasma temperatures, O abundance, and normalizations are free. \par

The first model utilizes a two-temperature \textsc{apec} model, assuming CIE for both components. The \textsc{apec} model is used to represent CIE thermal plasma components due to its success in describing a lower-temperature component of the Vela SNR spectrum \citep{2011PASJ...63S.827K}. This model fits the data adequately with a $\chi^{2}/dof = 1.24$. \par

The second model utilizes a two-component \textsc{pshock} model, which allows for the upper ionization timescales to be fitted to the spectrum as well as the standard parameters used in the first model. This model fits the data well with a $\chi^{2}/dof = 1.08$. When modeled with two NEI components, the hotter component of the Vela SNR spectrum is far from reaching collisional ionization equilibrium, having a fitted upper ionization timescale of $\tau_{u} = 0.34^{+0.08}_{-0.05} \times 10^{10}$ s cm$^{-3}$. The ionization timescale of the cooler component of the spectrum is $\tau_{u} = 5.0 \times 10^{13}$ s cm$^{-3}$ with a lower limit of $1.9 \times 10^{12}$ s cm$^{-3}$ and an undefined upper limit, thus the cooler component has reached collisional ionization equilibrium. As such, a third model was also tested to confirm the accuracy of the second model. \par

The third model utilizes a single-temperature \textsc{apec} model in conjunction with a single-component \textsc{pshock} model. The same parameters are free as were in the previous two models. This third model allows for comparison to the fitted upper ionization timescales of the two-component \textsc{pshock} model, on the assumption that the component in CIE is effectively modeled with an \textsc{apec} model. This model also fit the data well with a $\chi^{2}/dof = 1.08$. Given a $\tau_{u} = 5.0 \times 10^{13}$ s cm$^{-3}$ for one component of model 2, the congruence of models 2 and 3 was expected. \par

Both models accounting for NEI contributions in the Vela SNR fit the data over the $0.4-7$ keV band better, with a $\chi^{2}/dof = 1.08$ for both models, than the first model which has only CIE contributions, with a $\chi^{2}/dof = 1.24$. This is supported by an FTEST probability which indicates that models 2 and 3 fit the data better than model 1 does. Models 2 and 3 have FTEST probabilities of $6.5 \times 10^{-17}$ and $1.3 \times 10^{-17}$, respectively, as compared to model 1 in the $0.4-2.0$ keV energy band of interest. This supports the belief that the Vela SNR spectrum has one thermal component in CIE, and the other in NEI. Given the congruence of models 2 and 3, model 3 was used for its simplicity. A $\chi^{2}/dof$ value of $1.08$ is obtained for the fit, and the resulting spectrum is shown in Figure 3. \par

Fitted ionization timescales of the second and third models indicate that the spectral contributions from the Vela SNR are best fit with a two-temperature thermal plasma model with the cooler component in CIE with $kT_{1} = 0.19^{+0.01}_{-0.01}$ keV, and the hotter component affected by NEI contributions with $kT_{2} = 1.06^{+0.45}_{-0.27}$ keV. The O abundance is subsolar ($O/O_{\odot} = 0.36^{+0.09}_{-0.08}$), and all other abundances are set at solar values. The interstellar absorption column density is $N_{H} = 5.0^{+2.2}_{-2.1} \times 10^{20}$ cm$^{-2}$. \par

The Puppis A SNR contributions are best fit with a single-component NEI thermal plasma model with $kT = 0.86^{+0.06}_{-0.05}$ keV and $N_{H} = 2.1^{+0.54}_{-0.56} \times 10^{21}$ cm$^{-2}$ with solar abundances. The best-fit model parameters are given in Table 3, and the HS0009 fitted spectrum is shown in Figure 4.  \par

The response-weighting factors for each source are well constrained. The fitted Vela response-weighting factor is $0.05^{+0.01}_{-0.01}$, meaning that there are little spectral contributions from the Vela SNR in the HS0009 observation, as intended by the position of the pointing. We note that since only a small portion of the Vela SNR is in the HS0009 FOV, that portion's spectrum may differ from the spectrum of the entire SNR. However, since HS0009 is dominated by emission from the Puppis A SNR, this does not have a significant effect on our results. The fitted Puppis A response-weighting factor is $6.99^{+0.54}_{-0.26}$, meaning that the total count rate from Puppis A in HS0009 is $14.3^{+0.6}_{-1.0}$\% of the total count rate from Puppis A in HS0007. A separate check of this value is performed by simulating the detector response and FOV with the 12' ROSAT $0.56-1.21$ keV map of the Vela SNR and Puppis A SNR region (GPLANE: $l = 270 \text{, } b = 0$; R5). This calculation gives an expected value of 14.8\%, which is in agreement with the value fitted in XSPEC.

\begin{deluxetable}{lccc}[t]
\tablecaption{Best-fit parameters for the Vela SNR and Puppis A SNR}\label{tab:table}
\tablehead{
\colhead{Parameter\tablenotemark{c}}  & \colhead{} & \colhead{Vela SNR} & \colhead{Puppis A SNR} }
\startdata
 \multicolumn{4}{c}{\underline{\textsc{tbabs}}} \\
$N_{H}$ (10$^{21}$ cm$^{-2}$) & & $0.50^{+0.22}_{-0.21}$ & $2.05^{+0.54}_{-0.56}$ \\
\hline
 \multicolumn{4}{c}{\underline{\textsc{apec}}} \\
$kT$ (keV)&  & $0.19^{+0.01}_{-0.01}$ & -  \\
norm\tablenotemark{a}  & & $7.76^{+2.55}_{-2.44}$  & - \\
\hline
 \multicolumn{4}{c}{\underline{\textsc{pshock}}} \\
$kT$ (keV) & & $1.06^{+0.45}_{-0.27}$ & $0.86^{+0.06}_{-0.05}$ \\
$O/O\odot$ & & $0.36^{+0.09}_{-0.08}$ & - \\
$\tau_{u}$ ($10^{10}$ s cm$^{-3}$) & & $0.33^{+0.07}_{-0.05}$ & $15.2^{+4.8}_{-3.4}$ \\
norm\tablenotemark{a, b}  & & $2.07^{+0.92}_{-0.74}$ & $2.77^{+0.45}_{-0.35}$ \\
\hline
response-weighting factor & & $0.05^{+0.01}_{-0.01}$ & $6.99^{+0.54}_{-0.26}$ \\
$\chi^{2}$, $\chi^{2}/dof$ & & \multicolumn{2}{c}{$781.38$, $1.08$}
\enddata
\tablenotetext{a}{Normalizations are defined as $\frac{10^{-14}}{4\pi D^{2}} \int n_{e} n_{H} dV$ where $D$ is the distance to the source (cm), and $n_{e}$ and $n_{H}$ are the electron and H densities (cm$^{-3}$), respectively.}
\tablenotetext{b}{The normalization for the Puppis A SNR component is determined in the HS0009 observation, so it is scaled by the corresponding weighting factor to determine the normalization of the entire structure, which is given. }
\tablenotetext{c}{Errors are given for the 90\% confidence range for a single parameter. }
\end{deluxetable}

\begin{figure*}[t]
\epsscale{0.83}
\plotone{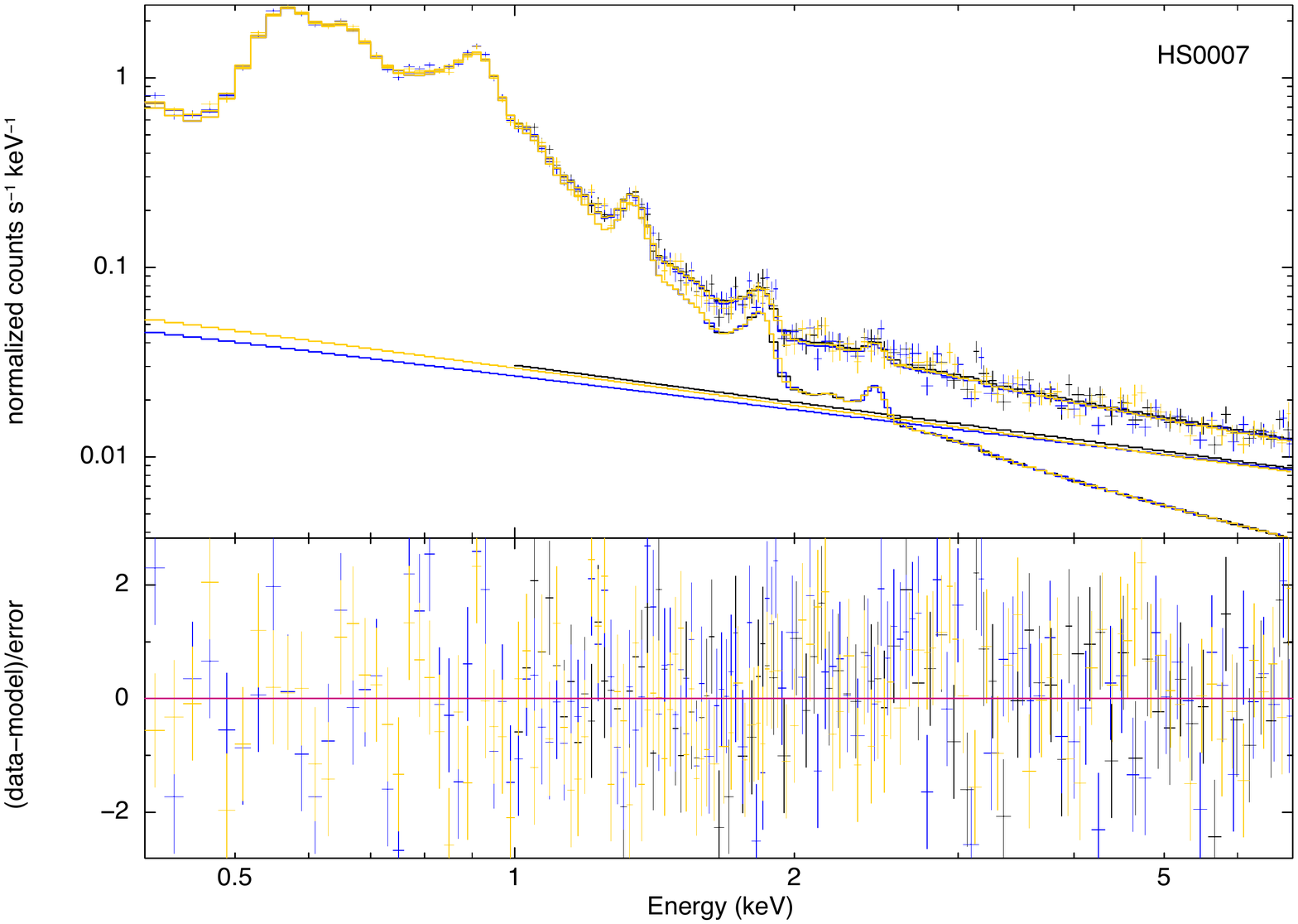}
\caption{HS0007 data and model from simultaneously fitting the HS0007 and HS0009 (Vela SNR and Puppis A SNR) spectra in XSPEC. The Vela SNR is modeled with a \textsc{tbabs} $\times$ (\textsc{vapec} $+$ \textsc{pshock}) model. The Puppis A SNR is modeled with a  \textsc{tbabs} $\times$ \textsc{pshock} model. $\chi^{2}/\text{dof} = 1.08$. Black: DPU 14; blue: DPU 54; gold: DPU 38. Linear components represent the instrumental background for each detector, model trends below the data represent the model for the astrophysical sources, and the trends through the data represent the sum of the astrophysical and instrumental model components.}
\plotone{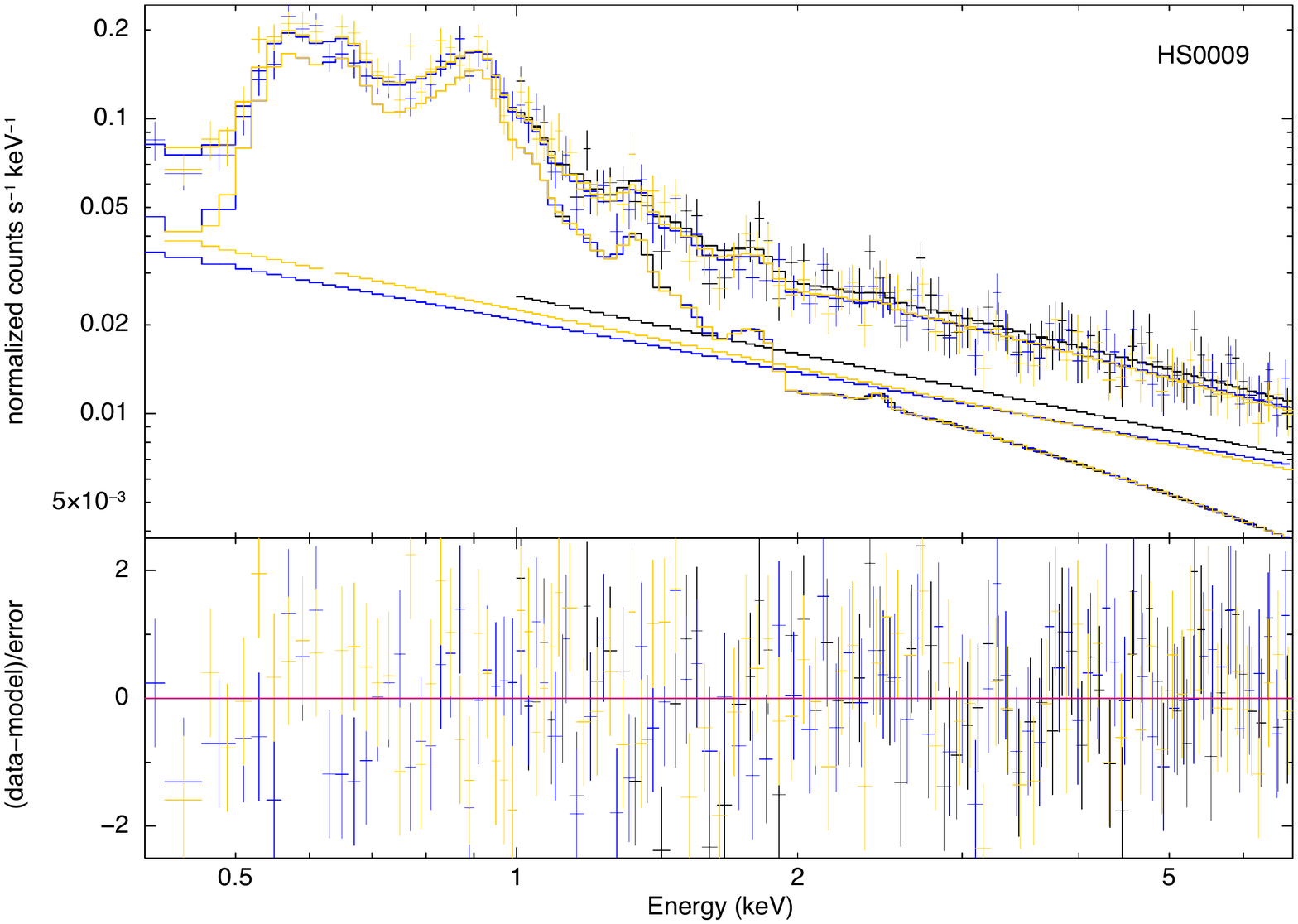}
\caption{HS0009 (Puppis A SNR) fitted spectrum with a  \textsc{tbabs} $\times$ \textsc{pshock} model and weighted contributions from the Vela SNR. $\chi^{2}/\text{dof} = 0.93$. Black: DPU 14; blue: DPU 54; gold: DPU 38. Linear components represent the instrumental background for each detector, model trends below the data represent the model for the astrophysical sources, and the trends through the data represent the sum of the astrophysical and instrumental model components.}
\end{figure*}

\section{Discussion}  \label{sec:disc}

\subsection{Vela SNR}

The ROSAT study of the entire Vela SNR by \citet{2000A&A...362.1083L} found that each region of the SNR was well fit by a two-temperature Raymond-Smith thermal plasma model in CIE with plasma temperatures of $kT_{1} \approx 0.09-0.25$ and $kT_{2} \approx 0.5-1.2$ keV and an interstellar absorption column density of $N_{H} = 5 \times 10^{19} - 6 \times 10^{20}$ cm$^{-2}$. However, due to the poorer spectral resolution of ROSAT ($\sim 5$ times poorer than HaloSat at 1.5 keV), NEI contributions could not be determined. With HaloSat, we find that the entire SNR may be adequately modeled with a similar two-temperature \textsc{apec} thermal plasma model in CIE, but is best-fit with a model which includes NEI effects in the hotter component. Plasma temperatures of $kT_{1} = 0.19^{+0.01}_{-0.01}$ keV and $kT_{2} = 1.06^{+0.45}_{-0.27}$ keV are constrained, as well as an interstellar absorption column density of $N_{H} = 5.0^{+2.2}_{-2.1} \times 10^{20}$ cm$^{-2}$. These fall within the ranges of temperatures and column densities found in the ROSAT study. \par

\citet{2000A&A...362.1083L} found the total unabsorbed X-ray flux and luminosity of the Vela SNR in the $0.1-2.5$ keV band to be $F_X = 2.9 \times 10^{-8}$ erg cm$^{-2}$ s$^{-1}$ and $L_X = 2.2 \times 10^{35}$ erg s$^{-1}$. The HaloSat spectra give a total unabsorbed X-ray flux and luminosity in the same $0.1-2.5$ keV band of $F_X = 4.8^{+1.5}_{-1.3} \times 10^{-8}$ erg cm$^{-2}$ s$^{-1}$ and $L_X = 3.6^{+1.2}_{-1.0} \times 10^{35}$ erg s$^{-1}$, where the uncertainties on the luminosity were estimated using only the uncertainties in the model normalization. The HaloSat unabsorbed flux and luminosity are believed to vary from the ROSAT values due to the larger interstellar absorption column density used in absorption-correction calculations. The HaloSat absorbed flux of the Vela SNR from $0.1-2.5$ keV is $F_X = 1.6^{+0.5}_{-0.4} \times 10^{-8}$ erg cm$^{-2}$ s$^{-1}$, in good agreement with the absorbed flux from ROSAT in the same band: $F_X = 1.3 \times 10^{-8}$ erg cm$^{-2}$ s$^{-1}$. Unlike the ROSAT study, the HaloSat observation allows for the total X-ray luminosity with errors to be determined from a single pointing. In the entire fitted energy band of $0.5-7$ keV, the total unabsorbed X-ray flux and luminosity are $F_X = 1.1^{+0.3}_{-0.3} \times 10^{-8}$ erg cm$^{-2}$ s$^{-1}$ and $L_X = 8.4^{+2.3}_{-2.0} \times 10^{34}$ erg s$^{-1}$. \par

Furthermore, \citet{2011A&A...525A.154S} notes that previous data have not been able to determine the X-ray luminosity of the hotter component of the Vela SNR spectrum with satisfactory accuracy. We find that for an assumed distance of $250$ pc in the same $0.5-7$ keV band, the total unabsorbed X-ray luminosities of each thermal component are $L_X = 4.4^{+1.4}_{-1.4} \times 10^{34}$ erg s$^{-1}$ for the cooler component and $L_X = 4.1^{+1.8}_{-1.5} \times 10^{34}$ erg s$^{-1}$ for the hotter component, where neither component dominates the total luminosity of the SNR. Thus, HaloSat data determines the X-ray luminosities of both thermal components with satisfactory accuracy. \par

The total X-ray luminosity of the Vela SNR in the $0.5-7$ keV band, $L_{\text{X}} = 8.4 \times 10^{34}$ erg s$^{-1}$, may be compared to the radio luminosity of the Vela $Y$ and Vela $Z$ regions at $1410$ MHz, $L_{\text{R}} = 1.5 \times 10^{34}$ erg s$^{-1}$ \citep{2001A&A...372..636A}. We find that $L_{\text{X}} / L_{\text{R}} = 1300$. Though the extent of radio emission in the $Y$ and $Z$ regions of the Vela SNR is smaller than the extent of the total X-ray emission, this result confirms the relative importance of X-ray contributions in the energetics of the Vela SNR.

A similar multi-ionization stage plasma result was achieved by \citet{2011PASJ...63S.827K} by modeling the soft X-ray spectrum of a region of the Vela SNR near Vela X with a hybrid two-temperature CIE/NEI plasma model. They found that, after including a non-thermal power-law component to represent harder spectral components associated with the Vela X PWN, the Vela SNR is represented by an absorbed two-temperature thermal plasma model where the cooler component is in CIE and the hotter component is in NEI (\textsc{tbabs} $\times$ (\textsc{apec} $+$ \textsc{vnei} or \textsc{pshock})). This model described lower plasma temperatures than both the ROSAT observation and this HaloSat observation with $kT_{1} \approx 0.09$ and $kT_{2} \approx 0.27$ keV. Their resulting interstellar absorption column density was also fixed at $N_H = 3 \times 10^{20}$ cm$^{-2}$. The lower fitted temperatures in their investigation may have to do with the implementation of a power-law component into the model to account for the Vela X PWN emission that was prominent in their observations at higher energies, which was not used in the HaloSat or ROSAT models. A further explanation may be how they focused on a small, specific region of the Vela SNR, whereas the ROSAT study and this analysis observe the entire SNR.  \par

\subsection{Puppis A SNR}

The Puppis A SNR is best fit by a single-component plane-parallel shocked plasma model in NEI (\textsc{pshock}), which is in agreement with previous studies of the SNR. \citet{2008ApJ...676..378H} used a similar model in fitting smaller regions across the Puppis A SNR, but found subsolar elemental abundances. By contrast, HaloSat observations indicate solar abundances. Using the \textsc{vpshock} model in place of the \textsc{pshock} model, subsolar abundances similar to those found by \citet{2008ApJ...676..378H} were tested with our best-fit model, but they did not improve the fit. It is noted by  \citet{2008ApJ...676..378H} that the parameters of any model may not be expected to be uniform throughout the entire remnant, as they are modeled in HaloSat observations. Despite this difference, \citet{2008ApJ...676..378H} did constrain plasma temperatures and interstellar absorption column densities in their various models which agree with the our results. Their temperatures ranged from $0.5-0.9$ keV, which contains the HaloSat value of $kT = 0.86^{+0.06}_{-0.05}$ keV, and their column densities ranged from $1.5-3.5 \times 10^{21}$ cm$^{-2}$, which are in agreement with the column density of $N_{H} = 2.1^{+0.54}_{-0.56} \times 10^{21}$ cm$^{-2}$ found in this analysis. \par

An analysis of the Puppis A SNR by \citet{2016A&A...590A..70L} also utilized an absorbed single-component \textsc{pshock} model to fit regions of the SNR. They observed no significant temperature fluctuations across the SNR, and found an average interstellar absorption column density of $N_{H} = 3.1 \times 10^{21}$ cm$^{-2}$ across the entire remnant, which ranged from approximately $N_{H} = 2.2-6.8 \times 10^{21}$ cm$^{-2}$ in selected regions. This range is also in agreement with the HaloSat value of $N_{H} = 2.1^{+0.54}_{-0.56} \times 10^{21}$ cm$^{-2}$ including errors. Plasma temperatures from \citet{2016A&A...590A..70L} ranged from approximately $0.45 - 0.62$ keV, which are lower than the fitted value of $kT = 0.86^{+0.06}_{-0.05}$ keV found in this analysis. In addition, their fitted elemental abundances are mostly subsolar. They suggest that the Puppis A SNR is a very structured SNR evolving in a complex environment, which may imply that single-temperature models may be oversimplified. However, the addition of a second temperature component to the Puppis A SNR model did not significantly improve the fit in this analysis. \par

We find a total unabsorbed X-ray flux and luminosity in the $0.5-7.0$ keV band for the Puppis A SNR of $F_X = 1.2 \pm 0.2 \times 10^{-8}$ erg cm$^{-2}$ s$^{-1}$ and $L_X = 6.7^{+1.1}_{-0.9} \times 10^{36}$ erg s$^{-1}$, for an assumed distance of 2.2 kpc. \citet{2013A&A...555A...9D} used \textit{Chandra} and \textit{XMM-Newton} observations to find an X-ray flux and luminosity in the $0.3-8$ keV band of $F_X = 2.2^{+1.4}_{-1.0} \times 10^{-8}$ erg cm$^{-2}$ s$^{-1}$ and $L_X = 1.2 \times 10^{37}$ erg s$^{-1}$. For the same $0.3-8$ keV energy band, we find a total unabsorbed X-ray flux and luminosity of $F_X = 1.5 \pm 0.2 \times 10^{-8}$ erg cm$^{-2}$ s$^{-1}$ and $L_X = 8.6^{+1.4}_{-1.1} \times 10^{36}$ erg s$^{-1}$. Our flux is consistent within errors to the value found by \citet{2013A&A...555A...9D}, sand our luminosity is well-constrained with reasonable errors. \par

The total X-ray luminosity of the Puppis A SNR in the $0.3-8$ keV band, $L_{\text{X}} = 8.6 \times 10^{36}$ erg s$^{-1}$, may be compared to the radio luminosity at $1425$ MHz, $L_{\text{R}} = 1.5 \times 10^{34}$ erg s$^{-1}$ \citep{2006A&A...459..535C, 2013A&A...555A...9D}, and the infrared luminosity, $L_{\text{IR}} = 5.4 \times 10^{37}$ erg s$^{-1}$ \citep{2010ApJ...725..585A}. We find that $L_{\text{X}} / L_{\text{R}} = 570$, and $L_{\text{IR}} / L_{\text{X}} = 6.2$. The $L_{\text{X}} / L_{\text{R}}$ and $L_{\text{IR}} / L_{\text{X}}$ ratios are similar to those found by \citet{2013A&A...555A...9D}, where $L_{\text{X}} / L_{\text{R}} = 800$ and $L_{\text{IR}} / L_{\text{X}} = 4.2$. These results confirm the substantial infrared and X-ray contributions in the Puppis A SNR spectrum as compared to radio contributions. \par

\section{Conclusions}  \label{sec:concl}
The only previous luminosity estimate for the entire Vela SNR was found by \citet{2000A&A...362.1083L}. They found that the spectrum could be modeled with a two-temperature thermal plasma model that was assumed to be in CIE due to the limited spectral resolution of ROSAT. We find that one hotter component in NEI is required, in addition to a cooler component in CIE to model the Vela SNR spectrum. We also find a higher X-ray luminosity in the $0.1-2.5$ keV energy band than \citet{2000A&A...362.1083L}. For the first time, the total X-ray luminosities in the $0.5-7$ keV energy band for each component of the Vela SNR spectrum are found:  $L_X = 4.4^{+1.4}_{-1.4} \times 10^{34}$ erg s$^{-1}$ for the cooler component and $L_X = 4.1^{+1.8}_{-1.5} \times 10^{34}$ erg s$^{-1}$ for the hotter component. This gives a total X-ray luminosity of  $L_{\text{X}} = 8.4 \times 10^{34}$ erg s$^{-1}$.

We confirm that the Puppis A SNR is best fit with a single component NEI plasma model. The total X-ray luminosity in the $0.5-7$ keV energy band is $L_X = 6.7^{+1.1}_{-0.9} \times 10^{36}$ erg s$^{-1}$. The fluxes and luminosities we find are similar to those found by \citet{2013A&A...555A...9D}, but with much smaller errors. \par

\section{Acknowledgements}  \label{sec:ackgnts}

The HaloSat mission is supported by NASA grant NNX15AU57G. This project/material is based upon work supported by the Iowa Space Grant Consortium under NASA Award No. NNX16AL88H. This research has made use of the MAXI data provided by RIKEN, JAXA and the MAXI team.
\facility{HaloSat}
\software{XSPEC \citep{1996ASPC..101...17A}}

\end{document}